\documentclass[aps,onecolumn,pra,superscriptaddress,groupedaddress]{revtex4}

\usepackage{amssymb} \usepackage{color,graphicx} \usepackage{amsmath}
\usepackage{amsbsy} \usepackage{amsthm}
\usepackage{float} \usepackage{braket}
\usepackage{placeins}
\usepackage[colorlinks=true,citecolor=blue,linkcolor=red,urlcolor=red]{hyperref}
\usepackage[sort&compress]{natbib}
\usepackage{comment}
\usepackage{array}
\newcolumntype{C}{>{$\displaystyle} c <{$}}

\newcommand{\bq}{\begin{equation}} \newcommand{\eq}{\end{equation}}
\newcommand{\bqali}{\bq\begin{aligned}}
\newcommand{\eqali}{\end{aligned}\eq}
\newcommand{\bqn}{\begin{equation*}}
\newcommand{\eqn}{\end{equation*}}
\newcommand\D{\operatorname{d}\!}

\renewcommand\k{{\bf k}}

\newcommand\rr{{\bf r}}
\newcommand\z{{\bf z}}

\newcommand\x{{\bf x}}

\newcommand\com[2]{[#1,#2]}

\newcommand\rC{r_\text{\tiny C}}

\begin{document}

\author{Emil Lenler-Eriksen}
\affiliation{Department of Physics and Astronomy, Aarhus University, Bygning 1520, Ny Munkegade, DK-8000 Aarhus C, Denmark}

\author{Michael Drewsen}
\affiliation{Department of Physics and Astronomy, Aarhus University, Bygning 1520, Ny Munkegade, DK-8000 Aarhus C, Denmark}

\author{Matteo Carlesso}
\email{matteo.carlesso@units.it}
\affiliation{Department of Physics, University of Trieste, Strada Costiera 11, 34151 Trieste, Italy}
\affiliation{Istituto Nazionale di Fisica Nucleare, Trieste Section, Via Valerio 2, 34127 Trieste, Italy}

\title{Testing Continuous Spontaneous Localization model with charged macromolecules}

\date{\today}

\begin{abstract}

In the last decade, a growing interest has been devoted to models of spontaneous collapse of the wavefunction, known also as collapse models. They coherently solve the well-known quantum measurement problem by suitably modifying the Schr\"odinger evolution. Quantum experiments are now finally within the reach of testing such models (and thus testing the limits of quantum theory). Here, we propose a method based on a two-ions confined in a linear Paul trap to possibly enhance the testing capabilities of such experiments. The combination of an atomic and a macromolecular ion provide a good match for the cooling of the motional degrees of freedom and a non-negligible insight in the collapse mechanism, respectively. 

\end{abstract}
\maketitle

\section{Introduction}
The validity of quantum theory at the microscopic scale has been verified to astonishing levels of precision. However, at the macroscopic scale, we perceive a fully classical world, where quantum superpositions are never observed although quantum mechanics does not pose any limitation.  Some modifications to the quantum theory, which connect these two limits, can  resolve the well-known quantum measurement problem. Among these, collapse models \cite{bassi2003dynamical} --- such as the Continuous Spontaneous Localisation (CSL) model \cite{pearle1989combining,ghirardi1990markov} --- introduce non-linear and stochastic (noisy) terms in the Schr\"odinger equation that lead to a spontaneous collapse of the wavefunction and coherently describe a quantum-to-classical transition when moving from micro to macro systems. The free phenomenological CSL parameters are the collapse rate $\lambda$ for a single nucleon and the correlation length $\rC$ of the collapse noise, respectively. A mass dependent amplification mechanism ensures that the more massive a system is, the faster it will collapse.  Testing the possible values of such parameters is equivalent to testing the limits of quantum mechanics. The current state-of-the-art in testing collapse models is achieved by recently developed non-interferometric experiments \cite{carlesso2022present}, which test collapse-induced  side-effects. Examples are diffusion, photon emission and  heating, and they can be measured with a plethora of different experimental setups \cite{carlesso2022present}.
Optomechanical tests placed among the tightest experimental upper bounds to date based on the quantification of  collapse heating and diffusion \cite{vinante2016upper,carlesso2016experimental, vinante2017improved, helou2017lisa, vinante2020narrowing, pontin2020ultranarrow, zheng2020room, komori2020attonewton}.

In this work, we propose an experimental  study of CSL through the detection of the motional heating of a two-ion system in a linear Paul trap, where the system is initially cooled to its ground state via Doppler and sideband laser cooling (see Ref.~\cite{eschner2003laser,ModeCoupling} for various ion cooling schemes). The sensitivity to CSL effects increases with the physical dimensions of the individual ions involved. Hence, in order to both ensure the initial ground state cooling and a high CSL sensitivity, we will consider a two-ion system consisting of a single atomic ion
and a charged macromolecule with an extended mass distribution. The former will be used for ground state sideband laser cooling of the common motional modes of the two-ion system, and any following heating of a specific mode  will be measured to quantify/bound potential CSL effects. {Notably, the use of molecules to test collapse models has been previously suggested \cite{bahrami2013collapse}, although envisioning an interferometric experiment.}

\section{The model} 
The CSL  master equation for the state $\hat \rho$ 
 reads \cite{carlesso2022present}
\begin{equation}\label{mastereq}
    \frac{\D \hat \rho}{\D t}=-\frac{i}{\hbar}\com{\hat H_0}{\hat \rho}-\frac{\lambda \rC^3}{2\pi^{3/2}m_0^2}
    \int\D \k\, e^{-k^2\rC^2} \com{\hat \mu_\k}{\com{\hat \mu_{-\k}}{\hat \rho}},
\end{equation}
where $m_0$ is a reference mass, taken equal to that of a nucleon, and $\hat \mu_\k$ is the Fourier transform of the mass density operator. Here, $\hat H_0$ is  the  free Hamiltonian   of the system, which is derived as follows.

The system consists of two ions of masses $m_1$ and $m_2$ and charges $q_1$ and $q_2$, respectively, placed in the same three-dimensional harmonic trap. The trap has a cylindrical symmetry along the $z$-axis. When the ion species are trapped independently, their motion along the $z$-axis corresponding to the frequencies $\omega_{i,z}$ with $i=1,2$, while $x$ and $y$ are the radial ones, corresponding to the frequencies $\omega_{i,r}$ with $i=1,2$. 
The total potential  reads
\bq
\hat V=\sum_{i=1,2}\tfrac12 m_i\left[\omega_{i,z}^2\hat z_i+\omega_{i,r}^2(\hat x_i^2+\hat y_i^2)\right]+\frac{q_1q_2}{4\pi\epsilon_0}\frac{1}{|\hat{\bf r}_1-\hat {\bf r}_2|},
\eq
where $\epsilon_0$ is the dielectric constant of vacuum. 
The Coulomb interaction  couples the ion motion. For sufficiently strong radial confinement, the ions will oscillate around the classical equilibrium positions ${\bf r}_{\text{eq},i}=(0,0,z_{\text{eq},i})$, where $z^3_{\text{eq},1}=\frac{q_1q_2}{4\pi\epsilon_0m_1\omega_1^2}({1+\tfrac{1}{Q}})^{-1}$, $z_{\text{eq},2}=-z_{\text{eq},1}/Q$ and $Q=q_2/q_1$,
with a motion that
can be described in terms of three in-phase and three out-of-phase modes. By Taylor-expanding the Coulomb interaction to the second order in the displacements, one obtains the  effective frequencies for the in-phase ($+$) and out-of-phase ($-$) motions along the axial and radial directions: $\omega_\text{Ax}^{\pm}$ and $\omega_\text{Rad}^{\pm}$.
They depend on the original axial and radial  frequencies $\omega_{i,z}$ and $\omega_{i,r}$ as well as the mass and charge   ratios $M = {m_2}/{m_1}$ and $ Q = {q_2}/{q_1}$ \cite{Supp}. We note that, due to the cylindrical symmetry of the trap, the radial frequencies have degeneracy two.
The in-phase and out-of-phase modes can then be described in terms of the creation and annihilation operators $(\hat a^\dag_{\pm},\hat a_{\pm})$, $(\hat b^\dag_{\pm},\hat b_{\pm})$, and $(\hat c^\dag_{\pm},\hat c_{\pm})$ where $a$ is for the axial modes and $(b,c)$ denote the $(x,y)$ radial modes respectively. In this paper we consider a cylindrically symmetrical trap, and as such the $x$ and $y$ directions can be treated identically. Thus, we can reduce the problem to a two-dimensional one. Hence, the free Hamiltonian of the two trapped ions can be expressed as
\bq
\hat H_0=\sum_{\nu\in\{+,-\}}\left[\hbar \omega_\text{Ax}^{\nu}(\hat a_\nu^\dag\hat a_\nu+\tfrac12)+\hbar \omega_\text{Rad}^{\nu}(\hat b_\nu^\dag\hat b_\nu+\tfrac12)\right].
\eq
One can express the displacement operators of the two ions with respect to the equilibrium positions in terms of $\hat a$, $\hat b$ and their hermitian conjugates. In particular, the displacement operator of the $i$-th ion along the $z$ direction with respect to $z_{\text{eq},i}$ is \cite{Supp}
\bq\label{eq.displacement.z}
\hat z_i=\sum_{\nu\in\{+,-\}}\sqrt{\frac{\hbar}{2 m_i \omega_\text{Ax}^\nu}}\alpha_{\nu,i}(\hat a^\dag_\nu+\hat a_\nu),
\eq
while the motion along $x$ (here the motion along $y$ is neglected) can be obtained from Eq.~\eqref{eq.displacement.z} by applying the substitutions 
 $\hat a_\nu\to\hat b_\nu$, $\omega_\text{Ax}^\nu\to\omega_\text{Rad}^\nu$ and $\alpha_{\nu,i}\to \beta_{\nu,i}$.  It should be noted that in the case of large charge-to-mass ratio mismatch between the two ions, each of the normal modes will be dominated by the motion of one ion, with the other only making a small contribution.

Finally, one computes the CSL-induced  effects by computing the collapse action on the single mode Hamiltonian $\hat h_\kappa$, {where $\kappa\in\{1,2,3,4\}$ labels the $\{\text{Ax}^+,\text{Ax}^-,\text{Rad}^+,\text{Rad}^-\}$ modes.} Specifically, from Eq.~\eqref{mastereq}, we compute the CSL-induced energy gain as $P_\kappa=\braket{\frac{\D \hat h_\kappa}{\D t}}$ of the mode $\kappa$.
By assuming that we can consider both ion species as point-like masses, one can substitute $\hat \mu_\k=\sum_im_ie^{-i\k(\rr_\text{eq,i}+\hat\rr_i)}$ and $\hat h_\kappa=\hbar \omega_\kappa(\hat u_\kappa^\dag\hat u_\kappa+\tfrac12)$, where $\hat u_\kappa$ can be either $\hat a_\kappa$ or $\hat b_\kappa$. Correspondingly, the commutators in Eq.~\eqref{mastereq} can be simplified, 
and integrating over $\k$, we find \cite{Supp}
\begin{widetext}
{\begin{equation}\label{Pkappa}
P_\kappa=\frac{\lambda\hbar^2}{4m_0^2\rC^2}\sum_{ij}\sqrt{m_i m_j}e^{-(\rr_{\text{eq},i}-\rr_{\text{eq},j})^2/4\rC^2}\left(1-\tfrac{(r_{\text{eq},\kappa,i}-r_{\text{eq},\kappa,j})^2}{2\rC^2}\right)\gamma_{\kappa,i}\gamma_{\kappa,j}^*,
\end{equation}}
\end{widetext}
{where $r_{\text{eq},1,i}=r_{\text{eq},2,i}=z_{\text{eq},i}$ and $r_{\text{eq},3,i}=r_{\text{eq},4,i}=0$.} In Appendix, we go beyond such an approximation,  considering the second ion as having an internal structure and a non-trivial non-deformable geometry.
The differences between the two approaches become visible only for values of $\rC$ smaller or comparable to the dimension of the macromolecule.

\textit{Competing effects. --}
Several  noises will be contributing to the overall heating mechanism. The strongest  will be due to electrical noises in the Paul trap, which will couple to the charges of the ions \cite{brownnutt2015ion}. They can be accounted for via standard perturbation theory with a perturbation  described by 
\begin{equation}
    \hat H_\text{\tiny E}(t)=-\sum_i q_i {\bf E}_t\hat \rr_i,
\end{equation}
where the sum runs over the ions, and ${\bf E}_t$ is a assumed spatially uniform stochastic electric field being characterised by zero mean $\mathbb E[{\bf E}_s]=0$ and whose correlations are assumed to be
\begin{equation}
    \mathbb E[E^{(k)}_sE^{(k')}_{s'}]=\frac{\delta_{k,k'}}{2\pi}\int_{-\infty}^{\infty}\D \omega\,S^{(k)}_\text{\tiny E}(\omega)e^{-i \omega(s-s')},
\end{equation}
where $S^{(k)}_\text{\tiny E}(\omega)$ is the corresponding power spectral density along the direction $k$. The consequent energy gain $P_\text{\tiny E}^{(\kappa)}$ of the mode $\kappa$ reads
\begin{equation}
    P_\text{\tiny E}^{(\kappa)}=\frac{\D}{\D t}\hbar \omega_\kappa\mathbb E\left[|c_\text{\tiny FI}|^2\right],
\end{equation}
where $c_\text{\tiny FI}$ is the transition amplitude  between the initial state (ground state of the $\kappa$ mode $\ket{0_\kappa}$) and final state (first excited state $\ket{1_\kappa}$), which reads
\begin{equation}
c_\text{\tiny FI}=-\frac{i}{\hbar}\int_0^t\D s\,\braket{1_\kappa|
e^{\tfrac{i}{\hbar}\hat H_0 s}\hat H_\text{\tiny E}(s)
e^{-\tfrac{i}{\hbar}\hat H_0 s}
|0_\kappa}.
\end{equation}
Following the detailed calculations reported in Appendix, and assuming that the electrical noises are spatially homogeneous, we find 
{\begin{equation}\label{Pelec}
P_{\text{\tiny E}}^{(\kappa)}\!=\!\sum_{ij}\frac{q_iq_j S_{\text{\tiny E},\kappa}(\omega_\kappa)}{2\sqrt{m_im_j}}\gamma_{\kappa,i}\gamma_{\kappa,j}^*,
\end{equation}
where $S_{\text{\tiny E},1}=S_{\text{\tiny E},2}=S_\text{\tiny E}^{(z)}$ and $S_{\text{\tiny E},3}=S_{\text{\tiny E},4}=S_\text{\tiny E}^{(x)}$, and}
which critically depends on the charges of the ions and their mass. By comparing the above expression with Eq.~\eqref{Pkappa}, one notice that \textit{i)} Eq.~\eqref{Pkappa} is independent from ions electrical charge, while  Eq.~\eqref{Pelec} goes as $\sim q_iq_j$; \textit{ii)} the mass dependence is strongly different: Eq.~\eqref{Pkappa} $\sim \sqrt{m_im_j}$ while Eq.~\eqref{Pelec} $\sim 1/\sqrt{m_im_j}$. From measurements with different ions, these different dependencies  make it possible to quantitatively differentiate motional heating due to electrical field fluctuations from other heating sources, including CSL.
Then, being able to perform a precise characterisation of electrical effects, one can  provide more stringent bounds CSL parameters.

\begin{figure}[t]
    \centering
    \includegraphics[width=0.6\linewidth]{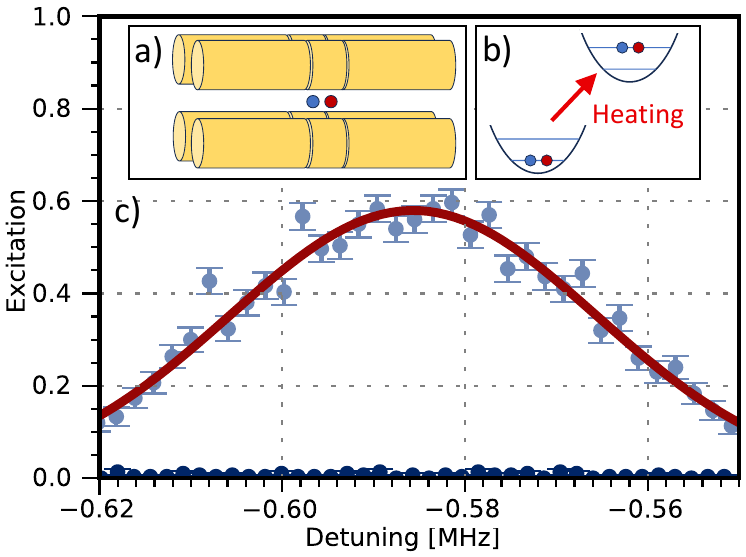}
    \caption{Experimental realisation. a)  Sketch  of a linear rf trap confining a laser-cooled atomic ion (blue disk), which sympathetically cools a charged macromolecule (red disk) through their mutual Coulomb interaction. A final sideband laser cooling step leads the system to its ground state. b) The possible collapse heating will eventually bring the system to its first excited motional state. c) The occurrence of an excitation can be verified using a red detuned laser. An excitation corresponds to a positive signal (light blue points and red fit), while no excitations lead to a null signal (dark blue points). The bottom graph is taken from \cite{poulsen2011sideband}. Details on these results can also be found in Ref. \cite{poulsen2012efficient}. }
    \label{Fig. Exp. method}
\end{figure}

\section{Experimental setup} 
A linear radio frequency (RF) trap \cite{drewsen2000harmonic} of rather large dimensions (distance between electrodes $\sim 5$\, mm, like that in Ref.~\cite{poulsen2012efficient}) will be applied. The trap will operate at cryogenic temperatures ($T\sim 4$\,K) to avoid strong influence of  heating mechanisms such as background gas collisions, thermally induced fluctuating electrical patch potentials on the electrodes, and scattering of blackbody radiation photons \cite{hansen2014efficient}. To achieve highly stable ions' trapping  in the RF trap (and have non-negligible cooling rates of at least one of the normal modes), we consider systems consisting of ions with similar mass-to-charge ratios. Consequently, we need a heavy atomic ion for ground state sideband laser cooling, and a massive potentially multiply charged macromolecule. Species such as Yb$^+$ or Hg$^+$ seem best suited based on their high masses, but the UV light needed for the laser cooling process of these ions may potentially be absorbed by the molecular ion and lead to unwanted internal heating or even fragmentation. With that in mind, the Ba$^+$ ion seems a more favorable choice. Any massive complex molecular ion which can be multiply charged and trapped could work. Here, we consider oligomers of porphyrin rings with each having two elementary charges,  a variety of such oligomers have been synthesized \cite{PorypherinSynthesis} and charged via {electrospray ion source \cite{Fenn,ProteinsESI,SuperChargerESI,ESIWilliams}}.

Once the  atomic ion is trapped  with the macromolecule, the following sequence is applied. 
First, the system is cooled to $\sim1 $mK by direct Doppler laser cooling of the atomic ion and indirect cooling of the molecular ion through the Coulomb interaction between the two ion species \cite{hojbjerre2008consecutive}. {To reach Doppler temperature of 1\,mK, it will eventually become necessary to couple the motional modes with high contribution from the molecule
to those with a high contribution from the Ba$^+$ ion (see for example \cite{ModeCoupling}), since only the latter will be cooled by Doppler cooling.} At this point the ions perform common-mode oscillations around their equilibrium positions indicated in Fig.~\ref{Fig. Exp. method}a. Next, ground state cooling of one of the common-modes is implemented through sideband laser cooling via addressing a narrow transition of the atomic ion on the so-called red sideband, i.e.~by exposing the ion to laser light with a frequency equal to the transition frequency minus the common-mode frequency. {Similarly, to reach the ground state of modes with low contribution from the Ba$^+$ ion, it will be necessary to perform mode coupling (see for example \cite{ModeCoupling}).} When the ground state is reached (Fig.~\ref{Fig. Exp. method}b, lower left), absorption on the red sideband ceases since any energy conserving excitation should lead to a reduction of a single common-mode quanta. This means, the atomic ion can no longer be brought into the long-lived excited state of the narrow transition, which can be confirmed experimentally with near $100\%$ efficiency through the application of a sequence of laser excitation pulses (dark blue points in Fig.~\ref{Fig. Exp. method} \cite{poulsen2012efficient,Supp}. After having checked in this way that the two-ion system is initialized in a common-mode ground state, the two-ion system is left unperturbed by laser light to heat by CSL mechanisms or other sources. After a time $\tau$, one checks if the ion system is still in the ground state. If an excitation of the common-mode  happened (Fig.~\ref{Fig. Exp. method}b, upper right panel), the laser excitation can now take place and can clearly be detected experimentally as indicated by the contrast between the dark and  light blue  points in Fig.~\ref{Fig. Exp. method}. Through repeated experiments with varying  $\tau$, one can eventually establish an experimental heating rate of the particular common-mode \cite{poulsen2012efficient}, which through comparisons with Eq.~\eqref{Pkappa}  can set an upper bound to CSL mechanisms, since any other unknown heating mechanisms would just lead to a faster  heating of the ion system.

By employing different molecular  species, for which CSL predicts 
different heating rates, one distinguishes the CSL heating from other sources. 
Better bounds on CSL can likewise be found by just changing the charge state of a specific molecule, since CSL heating is independent from the charge  [cf.~Eq.~\eqref{Pkappa}], contrary to that due to electrical field fluctuations [cf.~Eq.~\eqref{Pelec}]. {While heating rate in current ion trapping laboratories is at best $\sim 10^{-1} \sim 1$\,s$^{-1}$, and is dominated by various technical noise sources,  a much better noise control can be obtained by exploiting a special underground laboratory, which can be essentially free of interfering external electromagnetic field sources. We estimate that the corresponding ion heating rates from noise sources other than CSL can ultimately be as low as $10^{-5} \sim 10^{-6}$\,s$^{-1}$ for traps with electrode spacing of $\sim5$\,mm \cite{brownnutt2015ion}.}
{With recently designed DC-supplies, one can reach spectral voltage noise densities of at most $S_V(2\pi \times 100\,\text{kHz}) = 10^{-16}\,\text{V}^2/\text{Hz}$ (being limited by measurement noise) at a voltage of $10-20$\,V at room temperature. Then, by filtering these voltage supplies with cryogenically cooled RC filters, we expect that voltage noise densities below $S_V(2\pi \times 100\,\text{kHz}) = 10^{-24}\,\text{V}^2/\text{Hz}$ at temperatures of $T=2$\,K can be reached. 
At such low voltage noise densities, one needs also to account for Johnson noise. Again at 2\,K and an effective resistance of dielectric material of 0.1\,$\Omega$ at 100\,kHz, we calculate a Johnson noise density of $S_V^\text{Johnson}=10^{-23}$\,V$^2$/Hz, which will be the dominant contribution to the noise.
With our current trap design with an effective length scale of $D=56$\,mm, one can now estimate the heating rate by the formula \cite{brownnutt2015ion}:
\begin{equation}
    \frac{\D n}{\D t}=\frac{q^2}{4m \hbar \omega}S_\text{\tiny E}(\omega),
\end{equation}
where $S_\text{\tiny E}(\omega)=S_V/D^2$. With $q=24e$, $m=12000m_0$ and $\omega=2\pi \times 100\,$kHz, we find that $\D n/\D t=10^{-5}\,$s$^{-1}$.
}

\section{Results and Discussion} 
\begin{figure}[t]
    \centering
    \includegraphics[width=0.7\linewidth]{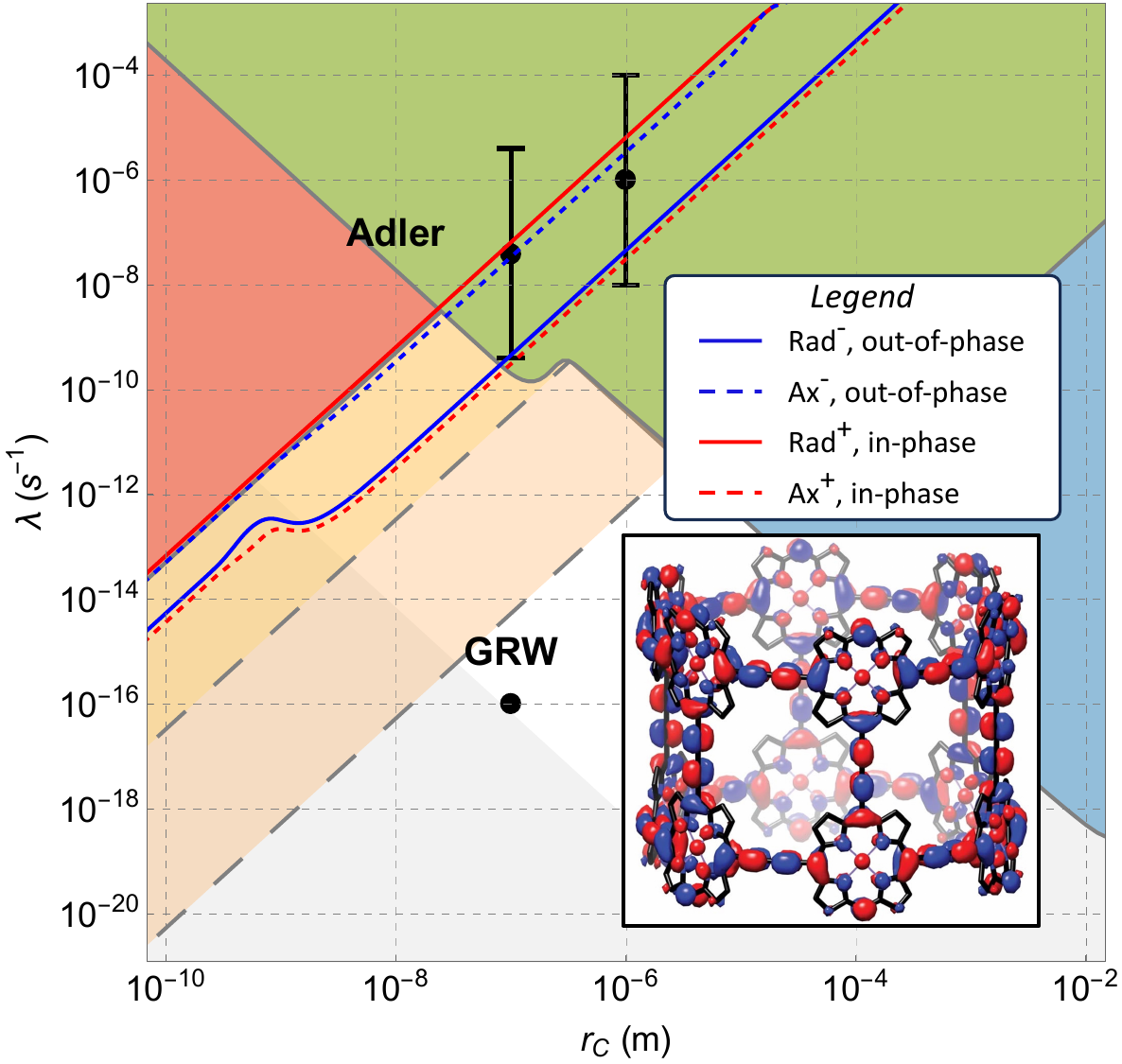}
    \caption{Bounds from our proposal with a $^{138}$Ba$^+$ atom and a 12-porphyrin molecule (see inset) aligned along the axial direction when compared with the state-of-art experimental bounds on the CSL parameters $\lambda$ and $\rC$. The expected energy gain of $P_\kappa=10^{-5}\times\hbar\omega_\kappa/\tau$ is considered.  For comparison, blue, green, red, yellow and orange shaded areas represent the experimentally excluded regions respectively from LISA Pathfinder \cite{carlesso2018non}, cantilever experiment \cite{vinante2020narrowing}, cold-atoms \cite{bilardello2016bounds}, phonons \cite{adler2019testing} and X-ray emission \cite{arnquist2022search}. {Phonons and X-rays emission bounds are strongly suppressed for coloured  CSL  with a cutoff frequency of $\Omega_\text{cut}\sim 10^{11}$\,s$^{-1}$, thus making our bounds the most stringent ones for $\rC\leq10^{-7}$\,m.} The gray region is theoretically excluded \cite{torovs2017colored}. Figure in inset from \cite{neuhaus2015molecular}.
    }
    \label{fig.csl-1}
\end{figure}

Here, we specifically consider the case of $^{138}$Ba$^+$ ($m_1=138 m_0$ and $q_1=1 e$) and a 12-porphyrin barrel-like structure (this is constructed with 2 rings made of 6 porphyrins, see inset in Fig.~\ref{fig.csl-1}) \cite{PorypherinSynthesis} with $m_2=8676m_0$ and $q_2=24|e|$ with $e$ being the elementary charge. 
Interestingly, the 12-porphyrin molecule can be used as a building block to construct more complex molecules, by stacking these barrel-like structures one on top of the other, and increase the mass while keeping the charge to mass ratio constant. The construction of porypherin nanotubes is discussed in Ref.~\cite{PorypherinSynthesis},{ which reports tubes consisting of up to 3 layers, i.e.~18 porphyrins. We underline that the employment of so constructed nanotubes is not the only available option. One could simply employ other natural or synthetised macromolecules providing the same or similar charge-to-mass ratio.}
The trap is prepared to have a single    
$^{138}$Ba$^+$ axial oscillation frequency of $\omega_{1,z}\simeq2\pi \times 100\,$\,kHz, while the radial frequency is set to $\omega_{1,r} \simeq 2\pi\times 242$ kHz is achieved.
The corresponding upper bounds {are obtained by equating the CSL heating rate to $P_\kappa=10^{-5}\times \hbar \omega_\kappa/\tau$ from the estimate of the electrical noise. The bounds  are presented in Fig.~\ref{fig.csl-1} with a comparison with the state of art \cite{carlesso2022present}}. These are computed going beyond the point-like approximation \cite{Supp}, with differences with respect to 
Eq.~\eqref{Pkappa} that can be seen in the bounds from the radial out-of-phase and axial in-phase modes only for values of $\rC$ below $10^{-9}\,$m.
\begin{figure}[t]
    \centering
        \includegraphics[width=0.5\linewidth]{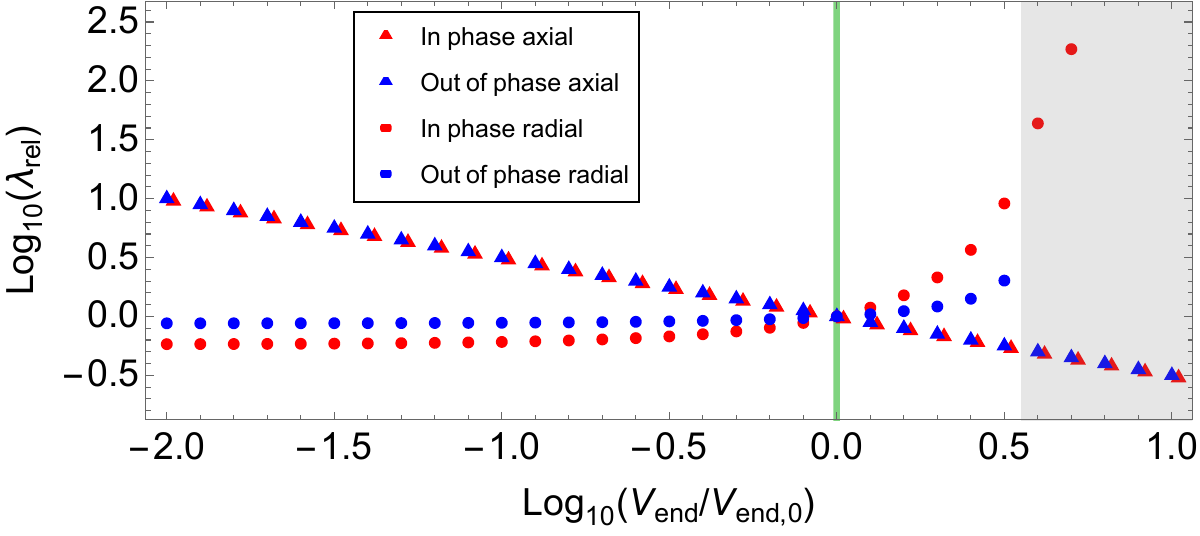}\includegraphics[width=0.5\linewidth]{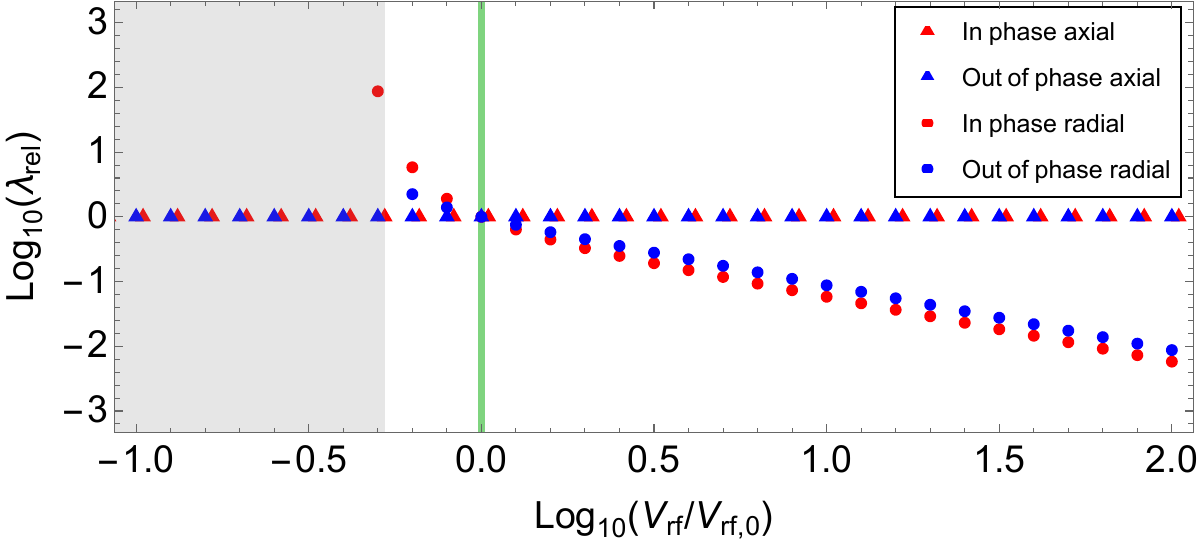}
    \caption{Relative improvements $\lambda_\text{rel}$ with respect to the results in Fig.~\ref{fig.csl-1} (highlighted by the green line) that can be achieved by changing: $V_\text{end}$ (left panel) and $V_\text{rf}$ (right panel).
        The reference values are $V_\text{end,0}=4.68\,$V and $V_\text{RF,0}=720.4\,$V. The grey-highlighted areas correspond to where the ions no longer are aligned along the $z$ axis, or no longer are stably trapped in the single ion case, as discussed in Appendix.
        }
    \label{fig.changefreq}
\end{figure}
Since the axial and radial modes' frequencies depend on   $V_\text{RF}$ and  $V_\text{end}$, they can be suitably tuned up to the stability limits of the linear Paul trap \cite{Supp}. The corresponding improvements in the CSL parameters' testing is  shown in Fig.~\ref{fig.changefreq}. Namely, a value of $\lambda_\text{rel}=10^N$ implies $N$ orders of magnitude improvement with respect to the bounds in Fig.~\ref{fig.csl-1}. {Overall, one can aim at a roughly two orders of magnitude improvement for the in-phase axial mode  with respect to what reported with the dashed red line in Fig.~\ref{fig.csl-1}  by setting $V_\text{end}=10^{-2}V_\text{end,0}$ and employing the equivalent of around 10 to 15 rings [see also Fig.~\ref{fig:LambDickes}]. Although this is not yet comparable with the bound from the X-ray emission, one needs to consider that the latter bound is not robust to coloured modifications to the CSL noise \cite{carlesso2022present,carlesso2018colored}, contrary to the bounds proposed here \cite{Supp} Indeed, for cosmologically motivated cut-off frequencies of $\Omega_\text{cut}\sim10^{12}\,$s$^{-1}$ \cite{ciufolini2010general}, the X-ray bound is completely washed out contrary to the  bounds from our proposal. Further, if one takes $\Omega_\text{cut}\sim10^{11}\,$s$^{-1}$ the blue continuous and red dashed bounds presented in Fig.~\ref{fig.csl-1}  become the strongest ones for $\rC\leq10^{-7}$\,m, as also the phonons' bound loses strength \cite{carlesso2018colored}.}

 An important experimental parameter is the Lamb-Dicke (LD) parameter $\eta_\kappa$, which describes the coupling strength of the light to the common motion of the ions. For the normal mode with frequency $\omega_\kappa$ and eigenvector $\gamma_{\kappa}$, it is defined as
 \begin{equation}
     \eta_{\kappa} = k\gamma_{\kappa,1}\sqrt{\frac{\hbar}{2m_1\omega_\kappa}}
 \end{equation}
 where $k$ is the wave vector of the laser light, and $\gamma_{\kappa,1}$ is the first entry of the normal mode vector $\gamma_\kappa$. 
 From an experimental perspective,  performing the readout on a specific mode $\kappa$ with near 100\% efficiency can be challenging for too small LD parameters and mode frequencies \cite{Supp}. Figure~\ref{fig:LambDickes} shows how $\eta$ changes for a different number of porypherin rings in the macromolecule. Here,  the axial out-of-phase mode and the radial in-phase mode seem to retain much higher LD parameters for all sizes of macromolecules. However, these are also the motional modes which are least sensitive to CSL heating, see Fig.~\ref{fig.csl-1}. {To perform readout on the modes with prohibitively low LD parameters we suggest transferring the phonon to a mode that is easier to perform readout on (i.e.~high LD parameter) by coupling the modes. This allows for combining the stronger CSL bounds gained from the modes with low LD parameters with the ease of readout on the modes with high LD parameters.}
 
 \begin{figure}[H]
    \centering
    \includegraphics[width=0.5\linewidth]{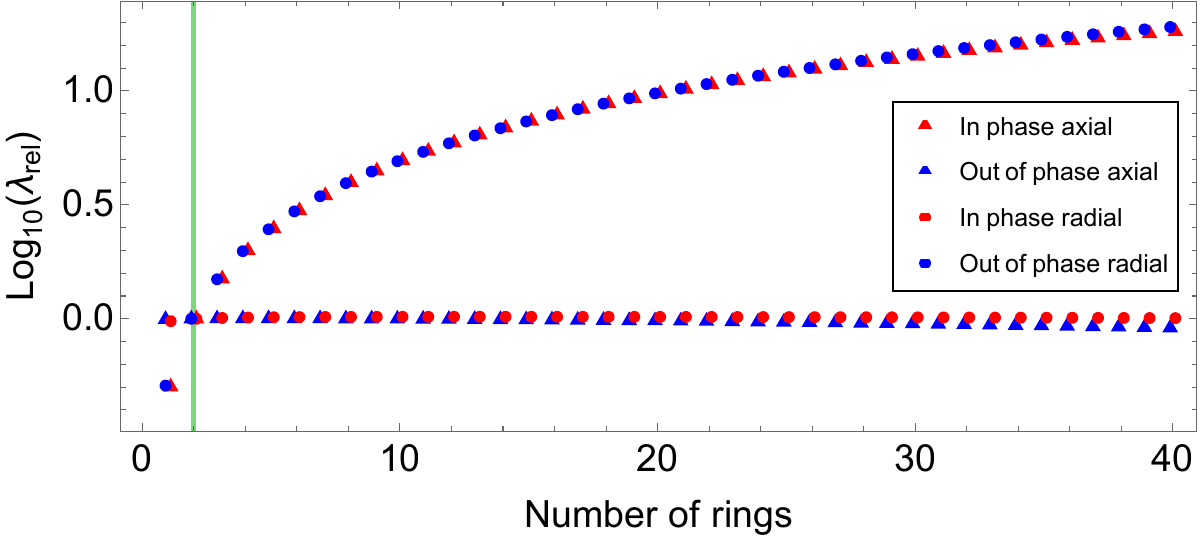}\includegraphics[width=0.5\linewidth]{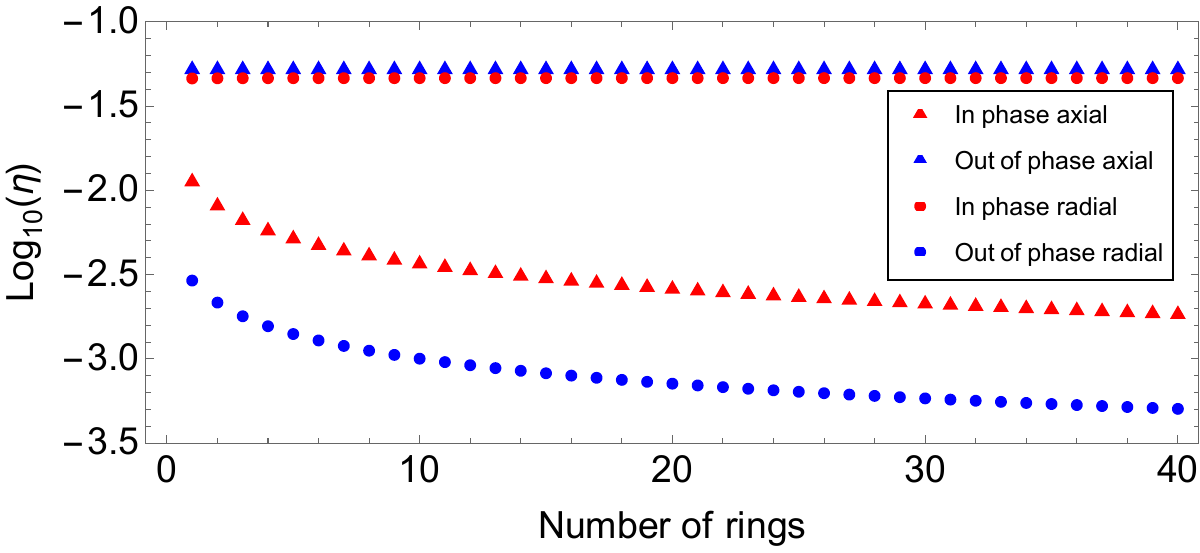}
    \caption{
    (Left panel) Relative improvements $\lambda_\text{rel}$ with respect to the results in Fig.~\ref{fig.csl-1} (highlighted by the green line) that can be achieved by changing the number of rings in the polypophyrin molecule, where each ring has 6 porphyrin molecules with a mass of $4338m_0$ and a charge of $12e$.
 (Right panel)   Lamb-Dicke parameter $\eta$ for axial and radial modes with 1762nm light interacting with the barium ion. Lamb Dicke parameters are calculated in a trap with $\omega_{1,z} = 2\pi\times 100$\,kHz and $\omega_{1,r} =2\pi\times242$\,kHz. We note that the in phase radial and out of phase axial modes have a very flat response to increasing number of porypherins, showing these could be good candidates for readout modes when working with very highly charged and massive ions. {Currently, it is realistically expected that a polypophyrin molecule with 10 rings can be constructed with state-of-the-art molecular engineering.}
 }
    \label{fig:LambDickes}
\end{figure}

Our proposal suggests the employment of a linear Paul trap with two ions for testing models of spontaneous collapse of the wavefunction. The two ions are chosen so that one allows for a large interaction with the collapse noise, while the other is employed for the initial cooling and the final measurement stages. The proposed method can be easily scaled to larger macromolecules, which will provide stronger insights to the collapse mechanism. Indeed, the limits of the proposal are only those related to the stability of the Paul trap, and the coherence time of the laser necessary for performing high efficiency readout of the common motion.\\

\section*{Acknowledgments}
MD and ELE appreciate support by the Danish National
Research Foundation through the Center of Excellence CCQ (Grant Agreement No. DNRF156) and the Independent Research
Fund Denmark – Natural Sciences via Grant No. DFF-1026-00293B.
MD and MC acknowledge the H2020 FET Project TEQ (grant n.~766900).
MC acknowledges the UK
EPSRC (Grant No.~EP/T028106/1), the EU EIC Pathfinder project QuCoM (10032223) and is  supported by the PNRR PE National Quantum Science and Technology Institute (PE0000023) and the University of Trieste (Microgrant LR 2/2011).

\appendix

\section{Derivation of the axial and radial motion of two ions in a linear Paul trap}
\subsection{The Linear Paul Trap}
\begin{figure}[b]
    \centering
    \includegraphics[width=0.6\linewidth]{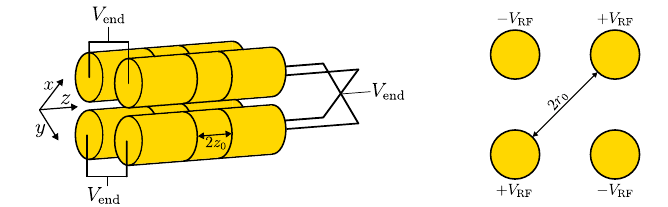}
    \caption{Schematic of the linear Paul trap, showing a 3D model of the trap (left) as well as a view along the $z$-axis (right). On each of the 4 rods a radiofrequency voltage of frequency $\Omega_\text{RF}$ and amplitude $V_\text{RF}$ is applied, while each of the 8 endcap electrodes also has a DC voltage $V_\text{end}$ applied. Furthermore we define $r_0$ as half the distance between diagonally opposed rods, while $z_0$ is half the distance between endcap electrodes. The values of the the parameters employed in the numerical analysis are presented in Table \ref{table:Trap}.}
    \label{fig:PaulTrap}
\end{figure}
\begin{figure}[h]
    \centering
    \includegraphics[width=0.5\linewidth]{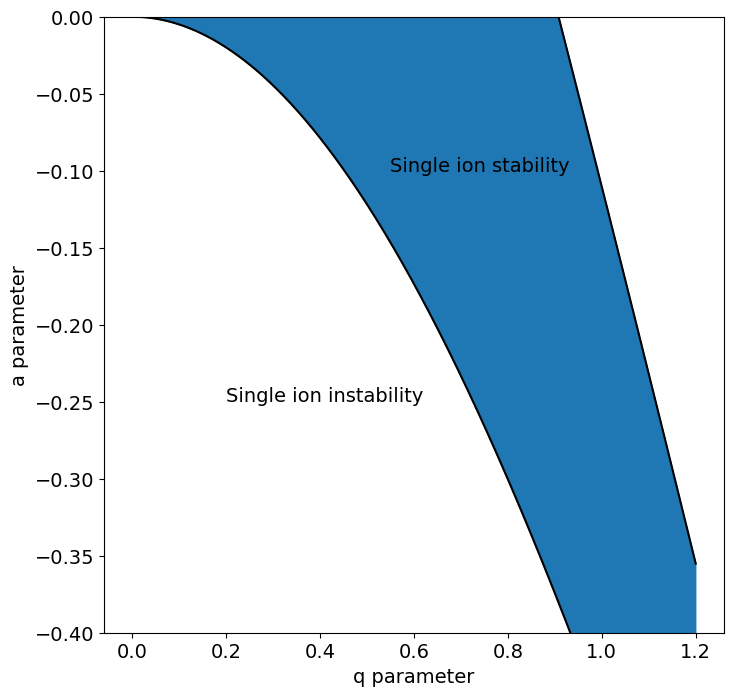}
    \caption{Stability diagram for the linear Paul trap. The black lines correspond to $a_0$ and $b_1$ as defined by Eq.~\eqref{eq: stablines} and the blue shaded area corresponds to the parameter space that gives stable trapping of positively charged ions. Note that the $b_1$ line is not yet visible in this part of the parameter space. The lack of stability for $a>0$ is due to the fact that, for the ion to be trapped along the $z$-direction, $V_\text{DC}$ must be positive and thus $a<0$.}
    \label{fig:paulstab}
\end{figure}

In this work, we focus on a linear Paul trap such as the one depicted in Fig.~\ref{fig:PaulTrap}. The trap consists of four rods, each of which is split in three electrodes as depicted on the figure. We define our coordinate system such that the $z$-axis runs along the center line of the linear Paul trap, while the $(x,y)$-axes are radial. We define $z_0$ as half the length of the center electrodes and $r_0$ as half the distance between diagonally opposed rods. A graphical depiction of $z_0$ and $r_0$ can be seen on Fig.~\ref{fig:PaulTrap}. When a DC voltage $V_\text{end}$ is applied to all electrodes located at the end of the rods (commonly referred to as endcap electrodes), the electrical potential in the center region of the trap will be
\begin{equation}
    \phi_\text{DC}(z) = \frac{\kappa V_\text{end}}{z_0^2}z^2,
\end{equation}
where $\kappa$ is a unitless constant determined by the specific geometry of the trap. Thus, a positively charged ion of mass $m$ and charge $q$, when placed at the center of the trap, undergoes to the following harmonic potential:
\begin{equation}
    V_\text{DC}(z) = \frac{q\kappa V_\text{end}}{z_0^2}z^2= \frac{1}{2}m\omega_z^2z^2\label{eq:harmonic_z},
\end{equation}
with the oscillation frequency $\omega_z$ of the ion along the $z$-axis being:
\begin{equation}
    \omega_z = \sqrt{\frac{2q\kappa V_\text{end}}{mz_0^2}}.\label{eq:omega_z}
\end{equation}
The applied voltage $V_\text{end}$ will, however, lead to an unconfining potential in the $(x,y)$-plane described by:
\begin{equation}
     \phi_\text{DC}(x,y) = -\frac{\kappa V_\text{end}}{2z_0^2}(x^2+y^2).
\end{equation}
To counteract this effect, we employ an RF voltage of frequency $\Omega_\text{RF}$, and peak-to-peak voltage $V_\text{RF}$ on all 4 rods, with adjacent rods having opposite phases as seen in Fig.~\ref{fig:PaulTrap}. The electrical potential in the radial plane may then, near the trap center, be written as \cite{Karin-Thesis}
\begin{equation}
    \phi(x,y,t) = \phi_\text{DC}(x,y)+\phi_\text{RF}(x,y,t) = -\frac{\kappa V_\text{end}}{2z_0^2}(x^2+y^2)-\frac{V_\text{RF}}{2r_0^2}(x^2-y^2)\cos{(\Omega_\text{RF}t)}.
\end{equation}
\begin{table}[t]
\centering
\begin{tabular}{ |c|c| } 
 \hline
 $\kappa$ & 0.248 \\ 
 \hline
 $\Omega_\text{RF}$& $2\pi\times 5.2$\,MHz\\ 
\hline
 $z_0$ & $2.03$\,mm\\ 
 \hline
 $r_0$ & $2.63$\,mm\\
 \hline
\end{tabular}
\caption{Values of the trap parameters employed in the numerical analysis: $\kappa$ is a unitless constant, defined by the specific geometry of the trap; $\Omega_\text{RF}$ is the RF frequency of the RF power supply; $z_0$ is half the distance between the endcap electrodes; and $r_0$ is half the distance between diagonally opposed electrodes. A graphical representation of $r_0$ and $z_0$ can be seen on Fig.~\ref{fig:PaulTrap}.}
\label{table:Trap}
\end{table}
In order to solve the equations of motion in the radial directions, it is useful to introduce three unitless parameters, which help find simplify the equations
\begin{equation}
    \tau = \frac{\Omega_\text{RF}t}{2},\quad a = -\frac{4q\kappa V_\text{end}}{mz_0^2\Omega_\text{RF}^2},\quad q_x = -q_y =  \frac{2qV_\text{RF}}{mr_0^2\Omega_\text{RF}^2}\label{eq:a and q}.
\end{equation}
We can use these parameters to put the equations of motion on a more compact form, finding \cite{Michael_EqMotion}
\begin{equation}
    \frac{\text{d}^2\rho}{\text{d}\tau^2} + (a-2q_\rho\cos{(2\tau)})\rho = 0,\quad \rho = x,y.
\end{equation}
This differential equation is known as the Mathieu equation \cite{Handbook_of_math} and has non-diverging (i.e.~stable) solutions when, for a given value of $q$, $a$ is found between the lines approximated by \cite{Handbook_of_math}
\begin{align}
    a_0(q) &\approx-\frac{1}{2}q^2+\frac{7}{128}q^4-\frac{29}{2304}q^6+\frac{68687}{18874368}q^8,\\
    b_1(q)&\approx1-q-\frac{1}{8}q^2+\frac{1}{64}q^3-\frac{1}{1536}q^4-\frac{11}{36864}q^5. \label{eq: stablines}
\end{align}
To trap positive ions in a stable way, it is necessary that we use a positive DC voltage, and as such it  necessary to have $a<0$. This, together with the requirement of a non-divergent solution, we get the stability diagram of the Paul trap reported Fig.~\ref{fig:paulstab}.
Experimentally, we often operate in the limit where $\vert a\vert , \vert q_{x,y}\vert \ll 1$. In this regime, we can obtain an approximate solution to the Mathieu equation
\begin{equation}
    \rho(t) = \rho_0\bigg[1-\frac{q_\rho}{2}\cos{(\Omega_\text{RF}t)}\bigg]\cos{(\omega_r t)} ,\quad \omega_r = \frac{\Omega_\text{RF}}{2}\sqrt{\frac{q_{\rho}^2}{2}+a},\quad \rho = x,y,\label{eq:radial_motion}
\end{equation}
We notice that, since $\vert q_\rho\vert \ll 1$, there is an oscillatory motion with small amplitude occurring at the RF frequency of the trap. This small, but rapid, motion is often referred to as micromotion. Further, such a micromotion is superimposed with another oscillating motion that is much higher in amplitude, but has a lower frequency as $\omega_r<\Omega_\text{RF}$, owing to the fact that $a$ and $q_\rho$ are both small.
If a time average is performed over the rapid micromotion, it will disappear leaving only simple harmonic motion in the radial plane, as if the particle was sitting in an effective potential (often called pseudopotential) given as
\begin{equation}
    V_r(x,y) = \frac{1}{2}m\omega_r^2(x^2+y^2).
\end{equation}
Combining this with Eq.~\eqref{eq:harmonic_z}, the total effective potential experienced by the ion can be described by a 3D harmonic potential, which exhibits a radial symmetry
\begin{equation}
    V_\text{tot}(x,y,z) = \frac{1}{2}m\bigg[\omega_z^2z^2+\omega_r^2(x^2+y^2)\bigg]
\end{equation}
and grants an harmonic motion along all three axes.
\subsection{Two ions in a linear Paul trap}
We now move on to the description of two ions in a linear Paul trap and their normal modes. We denote the two ions with the label  $i=1$ and  2 respectively, so that their mass and charge are respectively $m_i$ and $q_i$. Additionally we assume that each of the ions adhere to the stability requirements stated in the previous section. Finally we wish to consider the case where the ions align themselves along the $z$-axis. This requirement can be stated formally through the inequality
\begin{equation}
\frac{m_1m_2\omega_{1,z}^2\omega_{2,z}^2}{m_1\omega_{1,z}^2+m_2\omega_{2,z}^2}<\frac{m_1m_2\omega_{1,r}^2\omega_{2,r}^2}{m_1\omega_{1,r}^2+m_2\omega_{2,r}^2},\label{eq:z-aligned}
\end{equation}
where $\omega_{i,z}$ and $\omega_{i,r}$ are calculated as in Eq.~\eqref{eq:omega_z} and Eq.~\eqref{eq:radial_motion} using the mass and charge of ion $i$.

\begin{figure}[h]
    \centering
    \includegraphics[scale = 0.4]{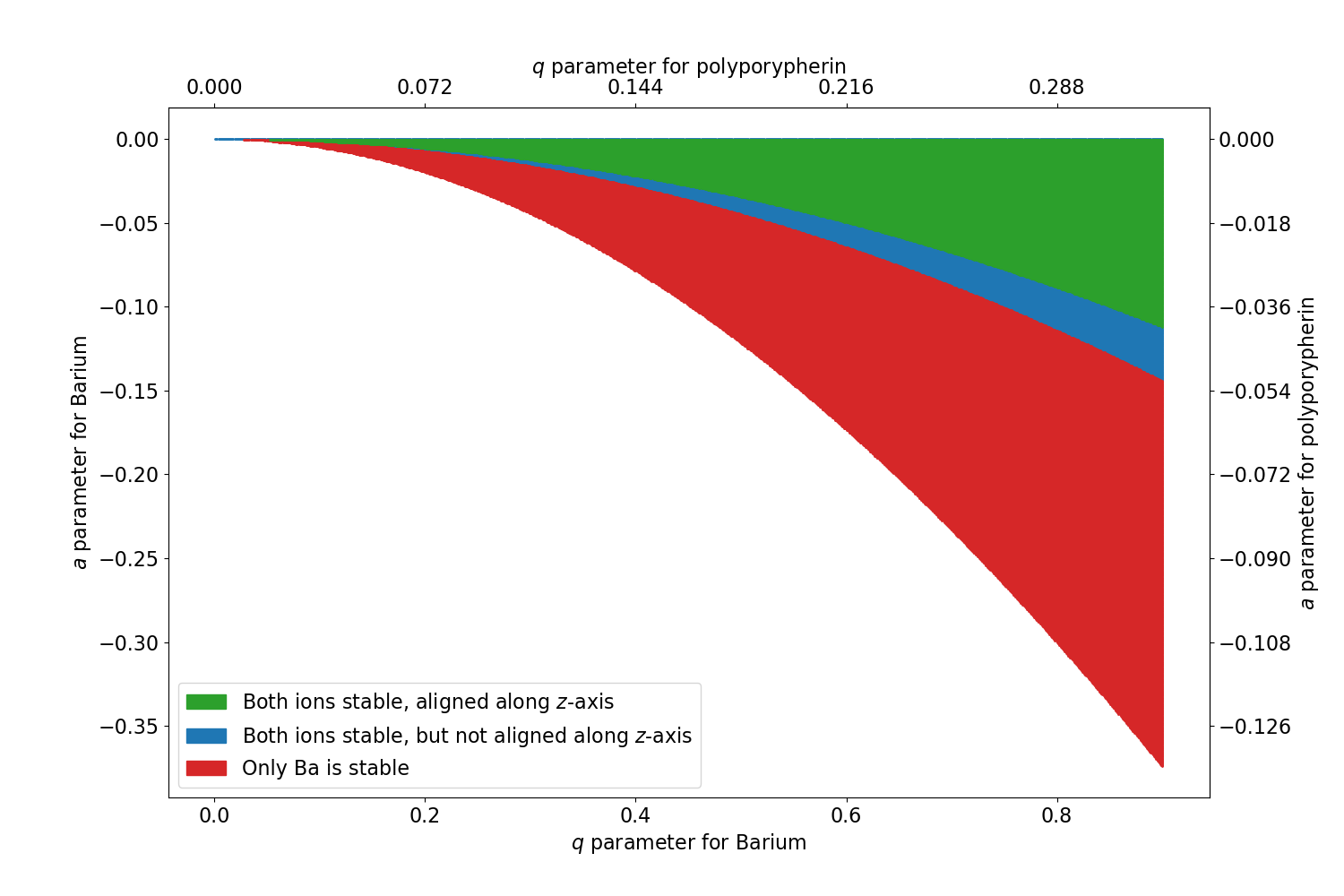}
    \caption{Stability region of the two-ion system consisting of a 24-charged polyporphyrin of mass $\sim9000$\,amu and a barium ion. The red area corresponds is where only the single ion stability criteria for the barium ion are met, the blue area is where the single ion stability criteria for the barium and the polyporypherin ion are met while the criterium for alignment along the $z$-axis is not, and finally the green area corresponds to fulfillment of single ion stability for both ions as well as alignment along the $z$-axis. It is to be noted that for the part of the $(q,a)$ parameter space shown in the figure there is no area where the polyporypherin would be the only ion to fulfill stable single ion trapping.}
    \label{fig:Stability_region}
\end{figure}

The requirement that both ions be stably trapped at the same time as well as lying along the $z$-axis puts additional constraints on the diagram seen in Fig.~\ref{fig:paulstab}. An effective stability diagram for the case of a two-ion system consisting of a barium ion and a +24-charged polyporypherin with molecular mass of 9000\,amu lying along the axis can be seen on Fig.~\ref{fig:Stability_region}, where it is visible that the range of permissible $a$ values is considerably narrower than the one shown on Fig.~\ref{fig:paulstab}. 

We begin the formal derivation of the normal modes of the two ions by determining their equilibrium positions. Working under the assumption that Eq.~\eqref{eq:z-aligned} is fulfilled, we may conclude that $x_{\text{eq},1}=x_{\text{eq},2}=y_{\text{eq},1}=y_{\text{eq},2}=0$ since the ions will be aligned along the $z$-axis. Letting $z_i$ be the $z$-cooridnate of ion $i$, and assuming without loss of generality that $z_1>z_2$ the potential energy of the system is

\begin{equation}\label{Potential1D}
    V(z_1,z_2) = \frac{1}{2}(m_1\omega_{1,z}^2z_1^2+m_2\omega_{2,z}^2z_2^2)+\frac{q_1q_2}{4\pi\epsilon_0}\frac{1}{ z_1-z_2}.
\end{equation}
To find the equilibrium positions $z_{\text{eq},i}$ of the two ions, we take the derivatives of  Eq.~\eqref{Potential1D} with respect to the different $z_i$ and set them equal to zero, when the system is in the equilibrium configuration.
\begin{subequations}
\begin{align}
    \frac{\partial V}{\partial z_1} = m_1\omega_{1,z}^2z_{\text{eq},1}-\frac{q_1q_2}{4\pi\epsilon_0}\frac{1}{(z_{\text{eq},1}-z_{\text{eq},2})^2}=0,\label{eq:dvdz1}\\
    \frac{\partial V}{\partial z_2} = m_2\omega_{2,z}^2z_{\text{eq},2}+\frac{q_1q_2}{4\pi\epsilon_0}\frac{1}{(z_{\text{eq},1}-z_{\text{eq},2})^2}=0.
\end{align}
\end{subequations}
Their sum gives
\begin{align}
    m_1\omega_{1,z}^2 z_{\text{eq},1}=-m_2\omega_{2,z}^2z_{\text{eq},2}\quad\longrightarrow  \quad z_{\text{eq},1} = -Q z_{\text{eq},2},
\end{align}
where we defined the charge ratio  $Q = {q_2}/{q_1}$. Replacing the expression for $z_{\text{eq},2}$ in Eq.~\eqref{eq:dvdz1} we find
\begin{equation}
m_1\omega_{1,z}^2z_{\text{eq},1} = \frac{q_1q_2}{4\pi\epsilon_0}\frac{1}{\big[z_{\text{eq},1}(1+\frac{1}{Q})\big]^2}\quad
    \longrightarrow \quad z_{\text{eq},1}^3 = \frac{q_1q_2}{4\pi\epsilon_0m_1\omega_{1,z}^2}\frac{1}{(1+\frac{1}{Q})^2}\label{eq:z(u)},
\end{equation}
which determines the equilibrium positions of both particles through known constants for a given experimental setup. From this it is clear that the equilibrium position of the ions does not depend on the mass since $\omega_{1,z} \propto \sqrt{m_1}^{-1}$.

\subsection{Multivariable Taylor expansion of the potential}\label{sec:Taylor}
To derive the normal modes along the axial and radial directions, we explicitly include the coordinates along the $x$ and $y$ axes. In particular, due to the degeneracy of the $x$ and $y$ motions imposed by the radial symmetry of effective potential, it is sufficient to consider only the motion along the $y$ axis. The motion along the $x$ axis is entirely identical.
The potential energy of the system is then described by
\begin{equation}
    V(z_1,y_1,z_2,y_2) = \frac{1}{2}(m_1\omega_{1,z}^2z_1^2+m_2\omega_{2,z}^2z_2^2+m_1\omega_{1,r}^2y_1^2+m_2\omega_{2,r}^2y_2^2)+\frac{q_1q_2}{4\pi\epsilon_0}\frac{1}{\sqrt{(z_1-z_2)^2+(y_1-y_2)^2}}.
\end{equation}
We now assume that the ions show only small oscillations around their equilibrium positions $\rr_{\text{eq},i}=(0,0,z_{\text{eq},i})$ and that these can be described in terms 
of time-dependent mass-weighted displacements $\zeta_j(t)/\sqrt{m_j}$, which are numbered with $j=1,\dots,4$. Namely, 
\begin{equation}\label{eq.def.xi}
    z_1(t) = z_{\text{eq},1}+\zeta_1(t)/\sqrt{m_1},\quad
    z_2(t) = z_{\text{eq},2}+\zeta_2(t)/\sqrt{m_2},\quad
    y_1(t) = \zeta_3(t)/\sqrt{m_1},\quad
    y_2(t) = \zeta_4(t)/\sqrt{m_2}.
\end{equation}
Now, we Taylor-expand the potential to the second order around the equilibrium positions, so that one has
\begin{equation}
    V\approx V_0+\frac{1}{2}\sum_{i,j=1}^4K_{ij}\zeta_i\zeta_j,
\end{equation}
where $V_0$ is a negligible constant energy offset and $K_{ij}=\left(\left.\frac{\partial^2V}{\partial r_i\partial r_j}\right|_\text{eq}\right)/\sqrt{m_im_j}$ with $\{r_1,r_2,r_3,r_4\}= \{z_1,z_2,y_1,y_2\}$ are the coefficients of the matrix
\begin{equation}\label{def.matrixK}
    K =
    \begin{pmatrix}
    K_{11} & K_{12} & 0 & 0\\
    K_{21} & K_{22} & 0 & 0\\
    0 & 0 & K_{33} & K_{34}\\
    0 & 0 & K_{43} & K_{44}\\
    \end{pmatrix}.
\end{equation}
The explicit form of the terms of $K$ are
\begin{equation}
\begin{aligned}
    K_{11} &= \omega_{1,z}^2 + \frac{q_1q_2}{4\pi\epsilon_0m_1}\frac{2}{\Delta z_{eq}^3}=\omega_{1,z}^2\left(1+\frac{2}{1+1/Q}\right),\\
K_{12} &= K_{21} =-\frac{q_1q_2}{4\pi\epsilon_0\sqrt{m_1m_2}}\frac{2}{\Delta z_{eq}^3}=-\frac{2\omega_{1,z}^2}{\sqrt{M}(1+1/Q)},\\
    K_{22} &= \omega_{2,z}^2 +\frac{q_1q_2}{4\pi\epsilon_0m_2}\frac{2}{\Delta z_{eq}^3}=\omega_{1,z}^2\frac{Q}{M}\left(1+\frac{2}{1+Q}\right),\\
K_{33} &= \omega_{1,r}^2-\frac{q_1q_2}{4\pi\epsilon_0m_1}\frac{1}{\Delta z_{eq}^3}=\omega_{1,r}^2 -\frac{\omega_{1,z}^2}{(1+1/Q)},\\
    K_{34} &= K_{43} = \frac{q_1q_2}{4\pi\epsilon_0\sqrt{m_1m_2}}\frac{1}{\Delta z_{eq}^3}=-\frac{1}{2}K_{12},\\
K_{44} &= \omega_{2,r}^2 - \frac{q_1q_2}{4\pi\epsilon_0m_2}\frac{1}{\Delta z_{eq}^3}=\omega_{2,r}^2 - \frac{\omega_{1,z}^2}{M(1+1/Q)},
\end{aligned}
\end{equation}
where $M=m_2/m_1$ and $\Delta z_{eq} = z_{1,eq}-z_{2,eq} =(1+Q)z_{1,eq}$.   The matrix $K$ in Eq.~\eqref{def.matrixK} is block diagonal, which means that the axial and radial motions are decoupled. Thus, the problem reduces to two eigen-problems of the $2\times2$ matrices:
\begin{equation}
    K_\text{Ax} = \begin{pmatrix}
    K_{11} & K_{12}\\
    K_{21} & K_{22}\\
    \end{pmatrix}
    ,\quad
    K_\text{Rad}=\begin{pmatrix}
    K_{33} & K_{34}\\
    K_{43} & K_{44}\\
    \end{pmatrix},
\end{equation}
which can be solved by requiring $\det(K_\text{Ax/Rad}-\lambda)=0$. The corresponding solutions are the in-phase ($+$) and out-of-phase ($-$) frequencies along the two directions, which are given by
\begin{equation}
\begin{aligned}
    &(\omega_\text{Ax}^{\pm})^2=\frac{K_{11}+K_{22}\mp\sqrt{(K_{11}-K_{22})^2+4K_{12}^2}}{2},\\
    &(\omega_\text{Rad}^{\mp})^2=\frac{K_{33}+K_{44}\mp\sqrt{(K_{33}-K_{44})^2+4K_{34}^2}}{2}.
\end{aligned}
\end{equation}
The axial frequencies explicitly read
\begin{equation}
    (\omega_\text{Ax}^\pm)^2=\frac{\omega_{1,z}^2}{2}\left(1+\frac{Q}{M}+\bigg(1+\frac{1}{M}\bigg)\frac{2Q}{1+Q}\mp\sqrt{\bigg(1-\frac{Q}{M}+\bigg(1-\frac{1}{M}\bigg)\frac{2Q}{1+Q}\bigg)^2+\frac{16Q^2}{M(1+Q)^2}}\right).
\end{equation}
 To find an expression for the radial frequencies, we define 
 \begin{align}
    \epsilon^2 = \frac{q_1V_\text{RF}^2z_0^2}{4\Omega_\text{RF}^2m_1r_0^4\kappa V_\text{end}}=\frac{1}{4}\Big\vert\frac{q_{1,x}^2}{a_1}\Big\vert=\frac{1}{4}\Big\vert\frac{q_{1,y}^2}{a_1}\Big\vert,
\end{align}
such that
\begin{equation}
\omega_{1,r} = \sqrt{\epsilon^2-\frac{1}{2}}\,\omega_{1,z}.
\end{equation}
Noting also that 
\begin{equation}
\omega_{2,r}=\sqrt{\frac{Q}{M}}\sqrt{\frac{\frac{Q}{M}\epsilon^2-\frac{1}{2}}{\epsilon^2-\frac{1}{2}}}\,\omega_{1,r},
\end{equation}
we find that
\begin{align}
    (\omega_\text{Rad}^{\pm})^2 = &\frac{\omega_{1,z}^2}{2}\left[\Big(\epsilon^2-\frac{1}{2}\Big)-\frac{1}{1+\frac{1}{Q}}+\chi^2\Big(\epsilon^2-\frac{1}{2}\Big)-\frac{1}{M(1+\frac{1}{Q})}\right.\nonumber\\
    &\mp\left.\sqrt{\Big[\Big(1-\chi^2\Big)\Big(\epsilon^2-\frac{1}{2}\Big)-\Big(1-\frac{1}{M}\Big)\frac{1}{1+\frac{1}{Q}}\Big]^2+\frac{4}{M(1+\frac{1}{Q})^2}}\right].
\end{align}
where we defined $\chi=\sqrt{\frac{Q}{M}}\sqrt{({\frac{Q}{M}\epsilon^2-\frac{1}{2}})/({\epsilon^2-\frac{1}{2}})}$. 

Once the axial and radial frequencies are obtained, we can compute the corresponding eigenmode vectors, which read
\begin{equation}
    \mathbf{\alpha}_{\pm} = \frac{1}{\sqrt{1+\tilde \alpha_{\pm}^2}}\begin{pmatrix}
    \tilde \alpha_{\pm}\\
    1
    \end{pmatrix},\quad\text{and}\quad
    \mathbf{\beta}_{\pm}=\frac{1}{\sqrt{1+\tilde \beta_{\pm}^2}}\begin{pmatrix}
    \tilde \beta_{\pm}\\1
    \end{pmatrix},
\end{equation}
with $\alpha$ denoting axial eigenvectors, $\beta$ denoting radial eigenvectors, and where we use $\tilde\alpha_{\pm}=[{(\omega_\text{Ax}^\pm)^2-K_{22}}]/{K_{12}}$ and $\tilde \beta_{\pm}=[{(\omega_\text{Rad}^\pm)^2-K_{44}}]/{K_{34}}$.

\subsection{Quantisation of the oscillatory modes}

The eigenmodes computed in Sec.~\ref{sec:Taylor} can be quantised via the standard approach, and they correspond to the single mode Hamiltonians
\begin{equation}
\hat h_\kappa=\hbar\omega_\kappa\hat u_\kappa^\dag\hat u_\kappa,
\end{equation}
where $\hat u^\dag_\kappa$ and $\hat u_\kappa$ are the creation and annihilation operators of the $\kappa$-th mode, where the label runs over the in-phase and out-of-phase modes along the axial and radial directions.
From $\hat u^\dag_\kappa$ and $\hat u_\kappa$, one can reconstruct the corresponding displacement operators \cite{Morigi2001}, which appear in  Eq.~\eqref{eq.def.xi}, by vector decomposition into the trap eigenmodes. This gives:
\begin{subequations}
\begin{align}
    &\frac{\hat{\zeta}_1}{\sqrt{m_1}} = \sqrt{\frac{\hbar}{2m_1\omega^+_\text{Ax}}}\mathbf{\alpha}_{+,1}(\hat{a}_++\hat{a}^\dagger_+)+\sqrt{\frac{\hbar}{2m_1\omega^-_\text{Ax}}}\mathbf{\alpha}_{-,1}(\hat{a}_-+\hat{a}_-^\dagger),\\
    &\frac{\hat{\zeta}_2}{\sqrt{m_2}}  = \sqrt{\frac{\hbar}{2m_2\omega^+_\text{Ax}}}\mathbf{\alpha}_{+,2}(\hat{a}_++\hat{a}^\dagger_+)+\sqrt{\frac{\hbar}{2m_2\omega^-_\text{Ax}}}\mathbf{\alpha}_{-,2}(\hat{a}_-+\hat{a}_-^\dagger),\\
    &\frac{\hat{\zeta}_3}{\sqrt{m_1}}  = \sqrt{\frac{\hbar}{2m_1\omega^+_\text{Rad}}}\mathbf{\beta}_{+,1}(\hat{b}_++\hat{b}^\dagger_+)+\sqrt{\frac{\hbar}{2m_1\omega^-_\text{Rad}}}\mathbf{\beta}_{-,1}(\hat{b}_-+\hat{b}_-^\dagger),\\
    &\frac{\hat{\zeta}_4}{\sqrt{m_2}}  = \sqrt{\frac{\hbar}{2m_2\omega^+_\text{Rad}}}\mathbf{\beta}_{+,2}(\hat{b}_++\hat{b}^\dagger_+)+\sqrt{\frac{\hbar}{2m_2\omega^-_\text{Rad}}}\mathbf{\beta}_{-,2}(\hat{b}_-+\hat{b}_-^\dagger),
\end{align}
\end{subequations}
where $\hat{a}_\pm$ and $\hat{b}_\pm$ are the annihilation operators for the axial and radial motion respectively, and $\alpha_{\nu,i}$ is the $i$-th entry of the eigenvector $\alpha_\nu$ and likewise for $\beta_{\nu,i}$.

\section{Light Matter Interaction}\label{sec:LMI}
To assess any motional heating of this system, we propose to perform an excitation on the so-called first red sideband (i.e.~using a laser with a frequency equal to the transition frequency minus the frequency of the addressed common motional mode $\nu$) with respect to the narrow transition. This can be also applied for sideband laser cooling, and thus used to initialize the two-ion system in the ground state of the specific mode.  With the laser being on resonance of this sideband transition, the sideband excitation probability of the $^{138}$Ba$^+ $ ion will be zero if the system is in the ground state of the specific motional mode, while the ion excitation probability in the case of a motional excitation to the first excited state will be given by
\begin{equation}
    P_\text{\tiny SB}(t) =\sin^2\bigg(\frac{\eta_{\nu,1}\Omega_0}{2}t\bigg)\label{eq:sideband},
\end{equation}  
where $\Omega_0$ is the carrier Rabi frequency when the two-ion system is in the motional ground state, $\eta_{\nu,1} = kw_{\nu,1}\sqrt{\frac{\hbar}{2m_1\omega_\nu}},$ with $ w\in\{\alpha_\pm,\beta_\pm\}$ is the Lamb Dicke parameter \cite{morigi2001two, wubbena2012sympathetic} describing the coupling strength of the laser light to the motional mode $\nu$, and $k$ is the wave vector of the laser light. For Eq.~\eqref{eq: rabi_osc} to be valid it requires $\eta_{\nu,1} \ll 1$. 
By choosing a readout pulse length $\tau_\text{read}$ such that
\begin{equation}\label{rabi.Lamb}
\eta\Omega_0\tau_\text{read} = \pi\to \tau_\text{read} = \frac{\pi}{\eta\Omega_0},
\end{equation}
we assure maximal electronic excitation of the $^{138}$Ba$^+ $ ion when the two-ion system is in the first excited motional state according to Eq.~\eqref{eq:sideband}.  
A laser detuned to the first red sideband will, however, be able to off-resonantly excite the $^{138}$Ba$^+ $ ion on the carrier transition when in the motional ground state. With the laser tuned to the first red sideband, it is at the same time always detuned by $\omega_\nu$ with respect to the carrier transition, which to first order leads to an electronic excitation probability given by 
\begin{equation}
    P_\text{c}(t) = \frac{\Omega_0^2}{\sqrt{\Omega_0^2+\omega_\nu^2}}\sin^2\left(\sqrt{\Omega_0^2+\omega_\nu^2}\frac{t}{2}\right)\label{eq: rabi_osc}.
\end{equation}
In order not to avoid “false positive” electronic excitations when the two-ion system is in the motional ground state, one will in a given implementation have to make sure that 
\begin{equation}\label{eq:off_res}
P_\text{\tiny SB}(\tau_\text{read}) \gg P_\text{c}(\tau_\text{read}),
\end{equation}
is fulfilled.
From Eq.~\eqref{eq:sideband} and Eq.~\eqref{eq: rabi_osc}, one sees that when the Lamb-Dicke parameter or the mode frequency become too small, it becomes challenging to fulfill inequality in Eq.~\eqref{eq:off_res} with respect to a reasonable readout pulse length.

\section{Beyond point-like approximation heating}

From Eq.~\eqref{mastereq} of the main text, one obtains the CSL-induced heating via 
\begin{equation}\label{d.hkappa}
    \frac{\D \hat h_\kappa}{\D t}=-\frac{\lambda \rC^3}{2\pi^{3/2}m_0^2}
    \int\D \k\, e^{-k^2\rC^2} \com{\hat \mu_\k}{\com{\hat \mu_{-\k}}{\hat h_\kappa}},
\end{equation}
where $\hat \mu_\k=\sum_im_ie^{-i\k(\rr_\text{eq,i}+\hat\rr_i)}$ and $\hat h_\kappa=\hbar \omega_\kappa(\hat u_\kappa^\dag\hat u_\kappa+\tfrac12)$.
Owning that
\begin{equation}\label{1stcomm}
{\com{\k\hat\rr_i}{\hat u_\kappa}=\sqrt{\frac{\hbar}{2m_i}} \frac{k_\kappa\gamma_{\kappa,i}}{\sqrt{\omega_\kappa}}  ,}
\end{equation}
is a c-number, and that $\com{\k\hat\rr_i}{\hat u_\kappa}=-\com{\k\hat\rr_i}{\hat u_\kappa^\dag}$, we can express
 \begin{equation}\label{2ndcomm}
\com{e^{-i\k\hat\rr_i}}{\hat u_\kappa^\dag\hat u_\kappa}=i\com{\k\hat\rr_i}{\hat u_\kappa^\dag}(\hat u_\kappa-\hat u_\kappa^\dag)e^{-i\k\hat\rr_i}.
 \end{equation}
{Here, we used the notation where $k_\kappa\in\{k_z,k_z,k_x,k_x\}$, $\omega_{\kappa}\in\{\omega_\text{Ax}^+,\omega_\text{Ax}^-,\omega_\text{Rad}^+,\omega_\text{Rad}^-\}$, and $\gamma_{\kappa,i}\in\{\alpha_{+,i},\alpha_{-,i},\beta_{+,i},\beta_{-,i}\}$.
}
Substituting Eq.~\eqref{2ndcomm}  in Eq.~\eqref{d.hkappa} and integrating over $\k$, we find that the CSL energy gain $P_\kappa=\braket{\frac{\D \hat h_\kappa}{\D t}}$ of the mode $\kappa$ is given by Eq.~\eqref{Pkappa} of the main text, which we report here
\begin{equation}\label{Pkappa_App}
  {P_\kappa=\frac{\lambda\hbar^2}{4m_0^2\rC^2}\sum_{ij}\sqrt{m_i m_j}e^{-(\rr_{\text{eq},i}-\rr_{\text{eq},j})^2/4\rC^2}\left(1-\tfrac{(\rr_{\text{eq},\kappa,i}-\rr_{\text{eq},\kappa,j})^2}{2\rC^2}\right)\gamma_{\kappa,i}\gamma_{\kappa,j}^*,}
\end{equation}
and assumes that both ions can be considered as point-like masses. 
Next, we go beyond such an approximation, considering  the second ion as having an internal structure and a non-trivial non-deformable geometry. It is now
constituted by a set of $N$ lighter masses $m_n^{(2)}$, such that $m_2=\sum_{n=1}^Nm_n^{(2)}$, whose positions are displaced from $(\rr_\text{eq,2}+\hat\rr_2)$ by the classical quantity $\rr_n^{(2)}$. The latter addition accounts for the geometry of the second ion and imposes the addition of a term similar to the round parenthesis in Eq.~\eqref{Pkappa_App} for the radial modes as well. Since we assume that the quantum fluctuations are the same for all constituents of the second ion, the relations in Eq.~\eqref{1stcomm} and Eq.~\eqref{2ndcomm} hold, while the mass dependence $\sqrt{m_i}$ explicitly appearing in Eq.~\eqref{Pkappa_App} are substituted by $m_n^{(i)}/\sqrt{m_2}$. The energy gain for the $\kappa$ mode  then reads

\begin{equation}\label{Pkappamulti}
  {  P_\kappa=\frac{\lambda\hbar^2}{4m_0^2\rC^2}\sum_{ij}\sum_{n,n'}\frac{m_n^{(i)}m_{n'}^{(j)}}{\sqrt{m_im_j}}e^{-(\rr_{\text{cl},n}^{(i)}-\rr_{\text{cl},n'}^{(j)})^2/4\rC^2}\left(1-\tfrac{(r_{\text{cl},\kappa,n}^{(i)}-r_{\text{cl},\kappa,n'}^{(j)})^2}{2\rC^2}\right)\gamma_{\kappa,i}\gamma_{\kappa,j}^*,}
\end{equation}
where $\rr_{\text{cl},n}^{(i)}=(0,0,z_{\text{eq,i}})+\rr_n^{(i)}$ and {$r_{\text{cl},1,n}^{(i)}=r_{\text{cl},2,n}^{(i)}=z_{n,i}$ and $r_{\text{eq},3,n}^{(i)}=r_{\text{eq},4,n}^{(i)}=x_{n,i}$}, while $m_n^{(1)}=m_1\delta_{n,1}$ and $\rr_n^{(1)}=0$ account for the fact that the first ion is made of a single point-like particle. As Eq.~\eqref{Pkappamulti} shows, a dependence on the alignment direction of the second ion arises at values of $\rC$ comparable with the dimensions of the ion.

\section{Electrical noise heating}

The heating action of an electrical noise can be accounted by adding to $\hat H_0$ the following stochastic Hamiltonian   \cite{poulsen2011sideband}
\begin{equation}
    \hat H_\text{\tiny E}(t)=-\sum_i q_i {\bf E}_t\hat \rr_i,
\end{equation}
where the sum runs over the particles, and ${\bf E}_t$ is a spatially uniform stochastic electric field being characterised by zero mean $\mathbb E[{\bf E}_s]=0$ and whose correlations are assumed to be
\begin{equation}
    \mathbb E[E^{(k)}_sE^{(k')}_{s'}]=\frac{\delta_{k,k'}}{2\pi}\int_{-\infty}^{\infty}\D \omega\,S^{(k)}_\text{\tiny E}(\omega)e^{-i \omega(s-s')},
\end{equation}
where $S^{(k)}_\text{\tiny E}(\omega)$ is the corresponding power spectral density along the direction $k$. The consequent energy gain $P_\text{\tiny E}$ can be then computed using perturbation theory and it reads
\begin{equation}
    P_\text{\tiny E}=\frac{\D}{\D t}\sum_\text{\tiny I}\sum_\text{\tiny F} \mathcal N_\text{\tiny I}\mathcal N_\text{\tiny F}(\epsilon_\text{\tiny F}-\epsilon_\text{\tiny I})\mathbb E\left[|c_\text{\tiny FI}|^2\right],
\end{equation}
where $\mathcal N_\text{\tiny I}$/$\mathcal N_\text{\tiny F}$ are, respectively, the probability of the initial/final state being $\ket{\text{I}}$/$\ket{\text{F}}$, whose corresponding energies are $\epsilon_\text{\tiny F}$ and $\epsilon_\text{\tiny I}$. Key to the calculation is the transition amplitude $c_\text{\tiny FI}$ between the initial and final states, which reads
\begin{equation}
c_\text{\tiny FI}=-\frac{i}{\hbar}\int_0^t\D s\,\braket{\text{F}|
e^{\tfrac{i}{\hbar}\hat H_0 s}\hat H_\text{\tiny E}(s)
e^{-\tfrac{i}{\hbar}\hat H_0 s}
|\text{I}}.
\end{equation}
In particular, here we are interested in the transition between the ground state $\ket{0_\kappa}$ (with $\mathcal N_\text{\tiny I}=1$) and the first excited state $\ket{1_\kappa}$ (with $\mathcal N_\text{\tiny F}=1$) of the $\kappa$-th mode of the two ions. Since these states are eigenstates of $\hat H_0$, the above expression becomes
\begin{equation}
c_\text{\tiny FI}=-\frac{i}{\hbar}\int_0^t\D s\,
e^{\tfrac{i}{\hbar}(\epsilon_\text{\tiny F}-\epsilon_\text{\tiny I}) s}
\braket{\text{F}|
\hat H_\text{\tiny E}(s)
|\text{I}}=
-\frac{i}{\hbar}\int_0^t\D s\,
e^{{i}{}\omega_\kappa s}
\braket{{1_\kappa}|
\hat H_\text{\tiny E}(s)
|{0_\kappa}}.
\end{equation}
By substituting the expressions of $\hat \rr_i$ in terms of the normal modes' ladder operators $\hat u_\kappa$ and $\hat u_\kappa$ [cf.~Eq.~\eqref{eq.displacement.z} of the main text], one finds that
{
\begin{equation}
    \braket{{1_\kappa}|
\hat H_\text{\tiny E}(s)
|{0_\kappa}}=\sum_iq_i\sqrt{\frac{\hbar}{2m_i}}
E_{\kappa,s}
\frac{\gamma_{\kappa,i}}{\sqrt{\omega_\kappa}}
,
\end{equation}
where $E_{1,s}=E_{2,s}=E^{(z)}_s$ and $E_{3,s}=E_{4,s}=E^{(x)}_s$.
}
Employing such a result, we see that $\mathbb E[|c_\text{\tiny FI}|^2]$ contains terms of the form
\begin{equation}\label{eq.deltat2}
    \int_0^t\D s\int_0^t\D s'\,e^{i 
 \omega_\kappa(s-s')}\mathbb E[E^{(k)}_sE^{(k')}_{s'}]=2\pi\delta_{k,k'}\int_{-\infty}^{\infty}\D\omega\,S_\text{\tiny E}^{(k)}(\omega)\left(\delta^{(t)}(\omega_\kappa-\omega)\right)^2,
\end{equation}
where we defined $    \delta^{(t)}(\omega_\kappa-\omega)$ such that
\begin{equation}
   \int_0^t\D s\,e^{i(\omega_\kappa-\omega)s}=2\pi e^{i(\omega_\kappa-\omega)t/2}\delta^{(t)}(\omega_\kappa-\omega), 
\end{equation}
for which the following relations hold
\begin{equation}
      \left(  \delta^{(t)}(\omega_\kappa-\omega)\right)^2\sim\frac{t}{2\pi}\delta^{(t)}(\omega_\kappa-\omega),\quad\text{and}\quad
      \lim_{t\to\infty}\delta^{(t)}(\omega_\kappa-\omega)=\delta(\omega_\kappa-\omega).
\end{equation}
Then, in the long-time limit, Eq.~\eqref{eq.deltat2} reduces to
\begin{equation}
      \int_0^t\D s\int_0^t\D s'\,e^{i\omega_\kappa(s-s')}\mathbb E[E^{(k)}_sE^{(k')}_{s'}]\sim\delta_{k,k'}\   t\  S_\text{\tiny E}^{(k)}(\omega_\kappa).
\end{equation}
It follows that the energy gain due to electrical noises on the $\kappa$ mode is given by
{
\begin{equation}
P_{\text{\tiny E}}^{(\kappa)}=\sum_{ij}\frac{q_iq_j S_{\text{\tiny E},\kappa}(\omega_\kappa)}{2\sqrt{m_im_j}}\gamma_{\kappa,i}\gamma_{\kappa,j}^*,
\end{equation}
where $S_{\text{\tiny E},1}=S_{\text{\tiny E},2}=S_\text{\tiny E}^{(z)}$ and $S_{\text{\tiny E},3}=S_{\text{\tiny E},4}=S_\text{\tiny E}^{(x)}$, and}
which corresponds to Eq.~\eqref{Pelec} of the main text.

\section{Optimising the collapse models bounds}

\begin{figure}[t]
    \centering
    \includegraphics[width=0.4\linewidth]{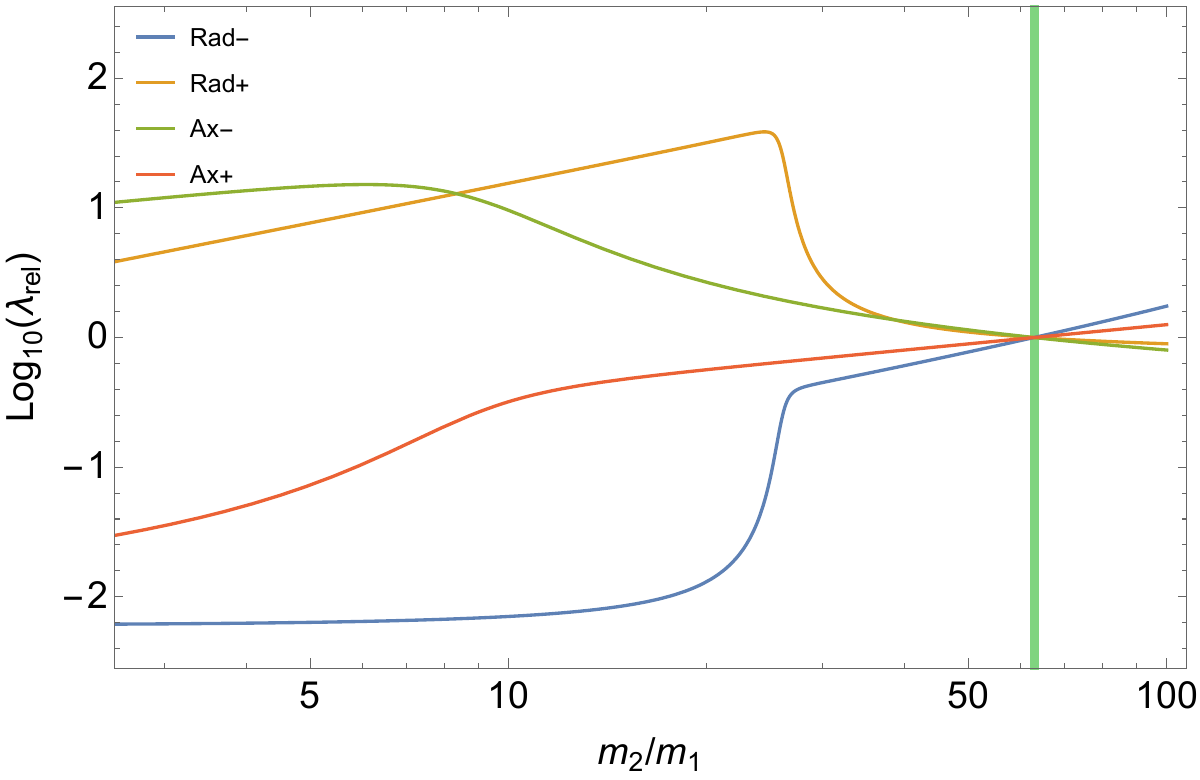}\includegraphics[width=0.4\linewidth]{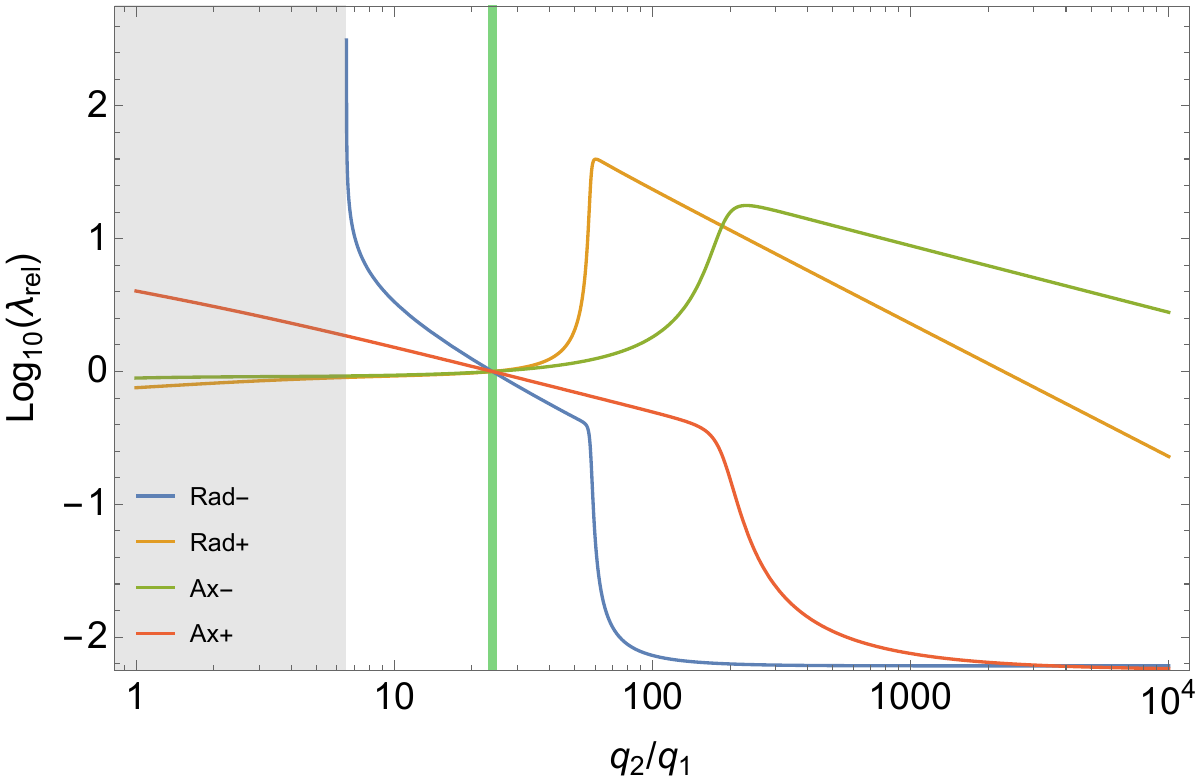}
    \caption{Relative improvements $\lambda_\text{rel}$ with respect to the results in Fig.~2 of the main text 
   (highlighted by the green line) that can be achieved by changing the mass (left panel) and the charge ratio (right panel) of the two ions. Results from all four modes are shown.  The the grey-highlighted area correspond to the Paul trap instability region.}
    \label{fig.change mass}
\end{figure}

\begin{figure}[t]
    \centering
        \includegraphics[width=0.9\linewidth]{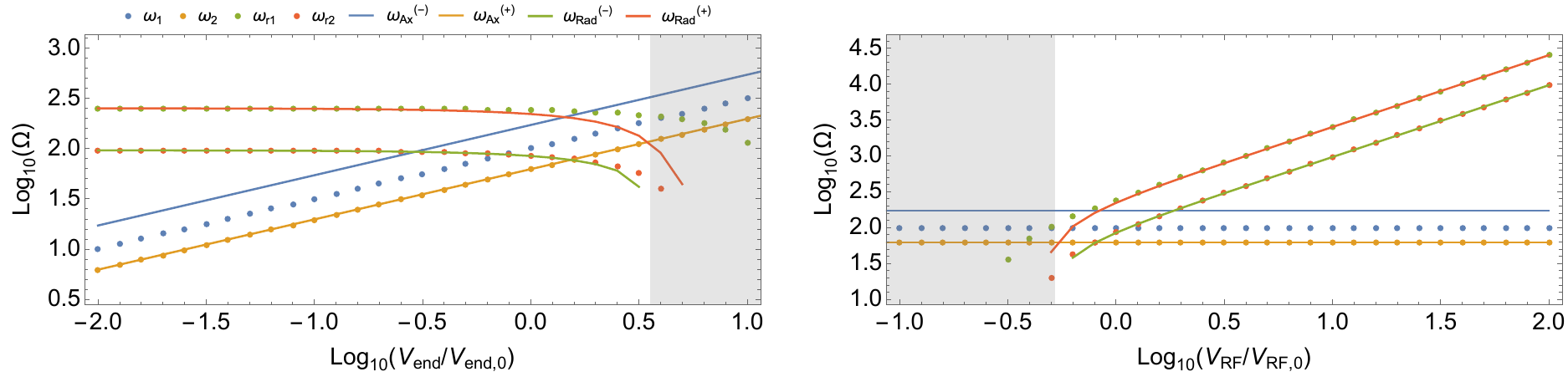}       
    \caption{Oscillations' and modes' frequency dependence from $V_\text{end}$ (left panel) and $V_\text{RF}$ (right panel). 
        The reference values are $V_\text{end,0}=4.68\,$V and $V_\text{RF,0}=720.4\,$V. The grey-highlighted areas correspond to the Paul trap instability region. }
    \label{fig.changefreq2}
\end{figure}

As it was indicated in the main text, one can optimise the bounds on the collapse models rate $\lambda$ by varying the mass and charge ratios as well as the number of rings $N_\text{\tiny rings}$ or the electric potentials in the trap. The results of such analysis are shown in Fig.~\ref{fig.change mass} where the mass and charge ratios are varied; and in Fig.~4 (top panel) of the main text where different values of $N_\text{\tiny rings}$ in the $6N_\text{\tiny rings}$-porphyrin molecule ion are considered. Figure~3 of the main text
 shows the change of the bounds when modifying the electric potentials $V_\text{end}$ and $V_\text{rf}$; the corresponding changes of the modes' frequencies are shown in Fig.~\ref{fig.changefreq2}.

\section{Colored modifications of CSL model}
{
The  CSL model can be mimicked 
with the addition of a stochastic potential to the
Schr\"odinger equation. Namely, such a potential reads \cite{carlesso2018colored}
\begin{equation}
\hat V_\text{\tiny CSL}(t)=-\frac{\hbar \sqrt{\lambda}}{m_0}\int\D\z\,\hat M(\z)\xi(\z,t),
\end{equation}
where $\hat M( \z)$ is the mass density operator and $\xi(\z,t)$ is a classical Gaussian noise characterised by
\begin{equation}
    \mathbb E[\xi(\z,t)]=0\,\quad \mathbb E[\xi(\z,t)\xi(\x,s)]=G(\z-\x)f(t-s).
\end{equation}
Here, $G(\x)$ is a normalised Gaussian of width $\rC$ and $f(t)$ is the time correlation of the collapse noise. When $f(t)=\delta(t)$ one has the standard CSL model. Conversely, when $f(t)$ differs from the Dirac delta function, one modifies the CSL model to its colored extension. Different choices of $f(t)$ provides different possible extensions. Nevertheless, they can  be  all characterised by a decay time-scale $\tau$ or equivalently by a cut-off  frequency $\Omega_\text{cut}$ of its Fourier transform. Such a frequency specifies that systems characterised by a frequency $\omega$ couple to the CSL noise only if $\omega<\Omega_\text{cut}$. 
Since the collapse should be universal,  a cosmological origin is reasonable. The cosmic microwave background (CMB) radiation and relic neutrino background provide indications of  $\Omega_\text{cut}\sim10^{12}\,$Hz \cite{carlesso2018colored}. For such a value, the bound on CSL from the X-ray emission is completely washed-out, while those presented in this work are robust to such a modification ($\omega\sim 2\pi\times 100$\,kHz).
}

\hfill

\bibliography{journal_version.bib}{}
\bibliographystyle{apsrev4-1}

\end{document}